\documentclass{article}
\usepackage{amssymb,amsmath,amscd, amsthm,stmaryrd,verbatim, mathrsfs}
\usepackage{color}

\newtheorem{Th}{Theorem}
\newtheorem{Def}{Definition}

\newcommand{\bb}{\mathbb}
\newcommand{\ms}{\mathscr}
\newcommand{\mr}{\mathrm}

\usepackage{hyperref}
\begin{document}

\title{A Generalization of the Kepler Problem}
\author{Guowu Meng\\
\small{\it Department of Mathematics, Hong Kong Univ. of Sci. and Tech.}\\
\small{\it Clear Water Bay, Kowloon, Hong Kong}\\
\small{Email: mameng@ust.hk} }

\maketitle

\begin{abstract}
We construct and analyze a generalization of the Kepler problem.
These generalized Kepler problems are parameterized by a triple $(D,
\kappa, \mu)$ where the dimension $D\ge 3$ is an integer, the
curvature $\kappa$ is a real number, the magnetic charge $\mu$ is a
half integer if $D$ is odd and is $0$ or $1/2$ if $D$ is even. The
key to construct these generalized Kepler problems is the
observation that the Young powers of the fundamental spinors on a
punctured space with cylindrical metric are the right analogues of
the Dirac monopoles.
\end{abstract}

\section{Introduction} The Kepler problem, owing to its
significant role in the major developments of physics in the last
three centuries, is a well-known scientific problem. It has been
known for quite a while that the Kepler problem and its beautiful
mathematical structure are not isolated: there are
deformations/generalizations along several directions. Firstly,
there is a generalization along the curvature direction
\cite{Sch40,Inf45}, i.e., deforming the configuration space from the
Euclidean space to the non-Euclidean space of constant curvature
$\kappa$; secondly, there is a generalization along the dimension
direction \cite{Sud65}, i.e., deforming the dimension (of the
configuration space) from three to a generic integer $D\ge 3$;
thirdly, there is a generalization along the magnetic charge
direction \cite{Z68,MC70}, i.e., deforming the magnetic charge (of
the nucleus of a hypothetic hydrogen atom) from $0$ to a generic
half integer $\mu$.

The Kepler problem has been generalized to the following cases in
the literature:
\begin{enumerate}
\item Case $\mu =0$, $\kappa \ge 0$, generic $D$, see
Refs. \cite{Hig79,Lee79},
\item Case $D =3$, $\kappa\ge 0$, generic $\mu$,
see Ref. \cite{Gri2000},
\item Case $D=3$, $\kappa<0$, generic $\mu$,
see Ref. \cite{NP2001} (where dimension five case was also
indicated),
\item Case $D=5$, $\kappa =0$, generic $\mu$,
see Ref. \cite{Iwai90}.

\end{enumerate} Note that, if $\mu \neq 0$, the generalized
Kepler problems are not expected to exist in dimensions other than
three and five. That is because, to get the Kepler problem with
magnetic charge in dimension other than three, the only method one
used so far is to apply the symmetry reduction to the isotropic
harmonic oscillator in a higher dimension. For example, the
isotropic harmonic oscillator in dimension eight admits a natural
$\mr {SU}(2)$-symmetry whose symmetry reduction to
$${\bb R}^8\setminus\{\vec 0\}/\mr{SU}(2)={\bb R}^5\setminus\{\vec
0\}$$ yields the five dimensional Kepler problem with magnetic
charge \cite{Iwai90}. Now the group $\mr{SU}(2)$ is the space of
unit quaternions with its action on $\bb R^8$ ($={\bb H}^2$) being
the scalar multiplication, and the set of quaternions, $\bb H$, is
an associative division algebra over $\bb R$. Since there are only
two nontrivial associative division algebras over $\bb R$ (the
complex numbers and the quaternions), the symmetry reduction method
can only produce the Kepler problems with nonzero magnetic charges
in dimensions three and five, which correspond to the complex
numbers and the quaternions respectively.

\vskip 10pt \emph{The main purpose here is to show that the Kepler
problem and its mathematical structure have a generalization beyond
what is currently known in the literature}. More explicitly, we
claim that there is a generalized Kepler problem for each triple
$(D, \mu, \kappa)$, where $D\ge 3$ is an integer, $\kappa$ is a real
number, and $\mu$ is a half integer if $D$ is odd and is $0$ or
$1/2$ if $D$ is even; moreover, these generalized Kepler problems
all share the characteristic mathematical beauty of the original
Kepler problem.

The key to our generalization in this generality lies in our recent
observation \cite{meng04} that the Young powers\footnote{Given an
irreducible representation $V$ of the orthogonal group, the $k$-th
Young power of $V$ is the irreducible representation whose highest
weight vector is equal to $k$ multiple of the highest weight vector
of $V$.} of the fundamental spinors on a punctured space with
cylindrical metric are the correct analogues\footnote{It is
interesting to note that, the magnetic charge, while it could be any
half integer in odd dimensions, it must be $0$ or $1/2$ in even
dimensions.} of the Dirac monopoles, see also Ref.
\cite{Cotaescu05}. With this observation in mind, it is basically
clear that how one can construct our generalized Kepler problems.
The details are given in the next section.

\vskip 10pt Our generalized Kepler problems are solved here by the
factorization method of Schr\"{o}dinger \cite{Sch40, Inf51}. In
principle, they can be solved by other methods such as the algebraic
method of Pauli \cite{Pauli26} and the path integral method
\cite{GS98, BIJ87}, though these other methods are technically more
difficult. Since the main purpose of this article is to inform the
community of the existence of the generalized Kepler problems in
this full generality, here we are contented with showing only the
factorization method in details and leave the solution by other
methods for the future.

\section{Generalized Kepler problems}
From the physics point of view, a generalized Kepler problem is
obtained from the Kepler problem by adding suitable background
gravity field plus magnetic field and making a suitable adjustment
to the scalar Coulomb potential so that the extra large hidden
symmetry is still preserved. The background gravity field is
introduced via the Riemannian metric for the sphere or
pseudo-sphere; and the background magnetic field is just the spin
connection of the cylindrical metric on the configuration space that
we have mentioned in the introduction. The configuration space is
the punctured Euclidean space if $\kappa =0$ or the punctured
pseudo-sphere if $\kappa <0$ or the sphere with two poles removed;
i.e., topologically it is either a punctured coordinate space if
$\kappa\ge 0$ or a punctured (open) disk if $\kappa < 0$. For
uniform treatment, we shall project the configuration space
isometrically onto the coordinate space. Note that, the cylindrical
metric is introduced \emph{only} for the purpose of introducing the
high dimensional analogue of Dirac's magnetic monopoles; we could
have introduced the background magnetic field by just writing down
the mysterious and hard-to-remember explicit formulae without
mentioning this true mathematical content. With this in mind, we are
now ready to give the detailed presentation of the generalized
Kepler problem.

\vskip 10pt

Let $\kappa$ be a real number. The \emph{configuration space} is a
$D$-dimensional Riemannian manifold $(Q_\kappa(D), ds^2_\kappa)$
with constant curvature $\kappa$, here $Q_\kappa(D)$ is $\bb
R^D-\{0\}$ if $\kappa\ge 0$ or $\{x\in \bb R^D\,|\, 0<|x|< {1\over
\sqrt {-\bar\kappa}}\}$  if $\kappa <0$, and
\begin{eqnarray}ds^2_\kappa={dx_0^2+\cdots+dx_{D-1}^2\over (1+\bar\kappa
r^2)^2},\end{eqnarray} where $r=|x|$ and $\bar\kappa={\kappa\over
4}$.

Let $ds^2={dx_0^2+\cdots+dx_{D-1}^2\over r^2}$. Just as in Ref.
\cite{meng05}, we observe that $(Q_\kappa(D), ds^2)$ is the product
of a line interval with the unit round sphere $\mr S^{D-1}$.  When
$D$ is odd, we let $\cal S_\pm$ be the positive/negative spinor
bundle of $(Q_\kappa(D), ds^2)$, and when $D$ is even, we let $\cal
S$ be the spinor bundle of $(Q_\kappa(D), ds^2)$. As we noted in
Ref. \cite{meng05} that these spinor bundles come with a natural
$\mr{SO}(D)$ invariant connection
--- the Levi-Civita spin connection of $(Q_\kappa(D), ds^2)$. As a
result, the Young product of $I$ copies of these bundles, denoted by
${\cal S}_+^I$, ${\cal S}_-^I$ (when $D$ is odd) and ${\cal S}^I$
(when $D$ is even) respectively, also comes with a natural
connection. For more details, the readers may consult Ref.
\cite{meng05}.

For the sake of notational sanity, from here on, when $D$ is odd and
$\mu$ is a half integer, we rewrite ${\cal S}_+^{2\mu}$ as ${\cal
S}^{2\mu}$ if $\mu\ge 0$ and rewrite ${\cal S}_-^{-2\mu}$ as ${\cal
S}^{2\mu}$ if $\mu\le 0$; moreover, we adopt this convention for
$\mu=0$: ${\cal S}^{0}$ is the product complex line bundle with the
product connection.

Introducing the pre-potential
\begin{eqnarray}
V_\kappa= -{1\over r}+{\bar\kappa}r
\end{eqnarray}
and factor
\begin{eqnarray}
\delta_\mu= \left\{\begin{matrix}(n-1)|\mu|+\mu^2 & \mbox{when $D=2n+1$} \\
(n-1)\mu & \mbox{when $D=2n$\;,}\end{matrix}\right.
\end{eqnarray}
we are then ready to make the following definition.
\begin{Def} Let $D\ge 3$ be an integer, $\kappa$ be a real number,
$\mu$ be a half integer if $D$ is odd and be $0$ or $1/2$ if $D$ is
even. The {\bf generalized Kepler problem} labeled by $(D, \kappa,
\mu)$ is defined to be a quantum mechanical system on $Q_\kappa(D)$
for which the wave-functions are sections of ${\cal S}^{2\mu}$, and
the hamiltonian is
\begin{eqnarray}
{\hat h}
 = -{1\over 2}\Delta_{\kappa}+{1\over 2}
\delta_\mu\left(V_\kappa^2+\kappa\right)+V_\kappa\, ,
\end{eqnarray}
where $\Delta_\kappa$ is the negative-definite Laplace operator on
$(Q_\kappa(D), ds_\kappa^2)$ twisted by ${\cal S}^{2\mu}$.
\end{Def}
Remark that this hamiltonian is an essentially self-adjoint operator
when some suitable ``boundary conditions" are imposed on the wave
functions, see Ref. \cite{Kato48} for the case $\mu=0$ and Ref.
\cite{BMS02} for the general case. However, we are not concerned
about this issue here because we can solve the eigenvalue problem
explicitly.

\section{Spectral analysis} We use the analytic method
pioneered by Schr\"{o}dinger \cite{Sch40} together with the textbook
knowledge of representation theory for compact Lie groups to obtain
the spectra for the bound states of our generalized Kepler problems.
As in Ref. \cite{meng05}, we assume that the small Greek letters,
$\alpha$, $\beta$, etc., run from $0$ to $D-1$ and the small Latin
letters, $a$, $b$, etc., run from $1$ to $D-1$. Under a local gauge
on $Q_\kappa(D)$ minus the negative $0$-th axis, the gauge
potential\footnote{We use the Einstein convention: the repeated
index is always summed over.}, $A=A_\alpha dx_\alpha$, is explicitly
calculated in Ref. \cite{meng05} as follows:
\begin{eqnarray}
A_0=0,\hskip 20 pt A_b=-{1\over r(r+x_0)}x_a\gamma_{ab}\,,
\end{eqnarray}
where $\gamma_{ab}={i\over 4}[\gamma_a,\gamma_b]$, and
$\gamma_a=i\vec e_a$ with $\vec e_a$ being the element in the
Clifford algebra that corresponds to the $a$-th standard coordinate
vector of $\bb R^{D-1}$.

Let $\nabla_\alpha=\partial_\alpha+iA_\alpha$. Then we can write
down the explicit local formula for $\Delta_\kappa$ as follows:
\begin{eqnarray}
\Delta_\kappa=(1+{\bar\kappa}r^2)^D\nabla_\alpha{1\over
(1+{\bar\kappa r^2})^{D-2}}\nabla_\alpha.
\end{eqnarray}
In view of that fact that $x_\alpha A_\alpha=0$, when written in
terms of the polar coordinates, the hamiltonian $\hat h$ then
becomes
\begin{eqnarray}
-{(1+\bar\kappa r^2)^D\over 2r^{D-1}}\partial_r(1+\bar\kappa
r^2)^{2-D} r^{D-1}\partial_r
 +{(1+\bar\kappa r^2)^2\over 2r^2}({1\over 2}\hat L_{\alpha\beta}^2-{\bar c}_2+\delta_\mu)
+V_\kappa,\nonumber
\end{eqnarray}
where
\begin{eqnarray}
\hat
L_{\alpha\beta}=-i(x_\alpha\nabla_\beta-x_\beta\nabla_\alpha)+r^2F_{\alpha\beta}
\end{eqnarray} and $\bar c_2$ is the value of the quadratic
Casimir operator of $\mr{so}(D-1)$ on the $2|\mu|$-th Young power
of the fundamental spin representation of $\mr{so}(D-1)$.

Let $L^2({\cal S}^{2\mu}|_{\mr{S}^{D-1}} )$ be the $L^2$-sections of
vector bundle ${\cal S}^{2\mu}$ restricted to the unit sphere
$\mr{S}^{D-1}$. From representation theory (for example, \S 129 of
Ref. \cite{DZ73}), we know that
\begin{eqnarray}
L^2( {\cal S}^{2\mu}|_{\mr{S}^{D-1}} )=\bigoplus_{l\ge 0}{\ms R}_l\,
,
\end{eqnarray} where, when $D$ is odd, ${\ms R}_l$ is the irreducible representation space of
$\mr{Spin}(D)$ having the highest weight equal to $(l+|\mu|, |\mu|,
\cdots,|\mu|)$; and when $D$ is even, ${\ms R}_l={\ms
R}_l^+\oplus{\ms R}_l^-$ with ${\ms R}_l^{\pm}$ being the
irreducible representation space of $\mr{Spin}(D)$ having the
highest weight equal to $(l+\mu, \mu, \cdots, \pm\mu)$. It is then
clear that the Hilbert spaces ${\ms H}$ of the bound states is a
subspace of
\begin{eqnarray}\label{predecompodd}
\bigoplus_{l\ge 0}{\cal H}_l\end{eqnarray} with ${\cal H}_l$ being a
subspace of $L^2(\mr I_\kappa, d\mu)\otimes {\ms R}_l$. Here $\mr
I_\kappa$ is the half line of the positive real numbers if
$\kappa\ge 0$ and is the open interval $(0, {1\over
\sqrt{-\bar\kappa}})$ if $\kappa<0$, and
\begin{eqnarray}d\mu=({r\over 1+\bar\kappa r^2})^D\,{dr\over
r}.\end{eqnarray}

On ${\cal H}_l$, ${1\over 2}L_{\alpha\beta}^2$ is a constant; in
fact, from the computation done in Ref. \cite{meng05}, we have
\begin{eqnarray}\label{c}
c:={1\over 2}L_{\alpha\beta}^2-{\bar
c}_2+\delta_\mu=m(m+1)-{(D-1)(D-3)\over 4}
\end{eqnarray} with $m=l+|\mu|+{D-3\over 2}$; so
the Schr\"{o}dinger equation becomes an ODE for the radial part:
\begin{eqnarray}
\left(-{(1+\bar\kappa r^2)^D\over
2r^{D-1}}\partial_r{r^{n-1}\over(1+\bar\kappa r^2)^{D-2}}
\partial_r+{(1+\bar\kappa r^2)c\over
2r^2}+V_\kappa-E_{kl}\right)R_{kl}=0\, ,
\end{eqnarray} where $E_{kl}$ is the $k$-th eigenvalue for a fixed $l$ and the additional label $k\ge 1$
is introduced for the purpose of listing the radial eigenfunctions,
just as in the Kepler problem. The further analysis follows the
factorization method in Ref. \cite{Inf51}. Making the transformation
\begin{eqnarray}
y=({r\over 1+\bar\kappa r^2})^{D-1\over 2}R_{kl}, \quad\quad
dx={1\over 1+\bar\kappa r^2}\,dr, \quad\quad \lambda =2E_{kl},
\end{eqnarray}
the preceding equation becomes

\begin{eqnarray}
{d^2y\over dx^2}+r(x,m)y+\lambda  y=0
\end{eqnarray}
with
\begin{eqnarray}
r(x,m) = {2(1-\bar\kappa r^2)\over r}-{(1+\bar\kappa
r^2)^2c+({D-1\over 2})^2(1-\bar\kappa r^2)^2-{D-1\over
2}(1+\bar\kappa r^2)^2\over r^2}.
\end{eqnarray}
By substituting the value for $c$ given in equation (\ref{c}), we
have
\begin{eqnarray}{r(x,m)\over -\kappa} =&
-{m(m+1)\over \sinh^2{\sqrt{-\kappa}x }}+ {1\over
\sqrt{-\bar\kappa}}\coth{\sqrt{-\kappa}x }-({D-1\over
2})^2&\hbox{for $\kappa<0$},\cr {r(x,m)\over \kappa} =&
-{m(m+1)\over \sin^2{\sqrt{\kappa}x}}+ {1\over
\sqrt{\bar\kappa}}\cot{\sqrt{\kappa}x}+ ({D-1\over 2})^2&\hbox{for
$\kappa>0$}, \cr r(x,m) =& {2\over r}-{m(m+1)\over r^2}&\hbox{for
$\kappa=0$}.
\end{eqnarray}

Let $I_0(\kappa)$ be $\infty$ if $\kappa\ge 0$ or be the integer
part of $\left((-\kappa)^{-{1\over 4}}-|\mu|-{D-1\over 2}\right)$ if
$\kappa <0$. Using the factorization method \cite{Inf51} plus
theorems 2 and 3 in \S 129 of Ref. \cite{DZ73}, one can arrive at
the following theorem for the generalized Kepler problems.
\begin{Th}  For the generalized Kepler problem labeled by $(D, \kappa, \mu)$,
assume $I_0(\kappa)\ge 0$, then the following statements are true:

1) The discrete energy spectra are \begin{eqnarray}\label{spec}
E_I=-{1\over 2(I+{D-1\over 2}+|\mu|)^2}+{(I+{D-1\over
2}+|\mu|)^2-({D-1\over 2})^2\over 2}\kappa\, ,
\end{eqnarray}
where $I$ is a non-negative integer and is less than or equal to
$I_0(\kappa)$;

2) The Hilbert space $\ms H$ of the bound states admits a linear
$\mr{Spin}(D+1)$-action under which there is a
decomposition\footnote{This follows from a fact in representation
theory: the direct sum of certain suitable irreducible
representation spaces of ${\rm Spin}(D)$ forms an irreducible
representation for ${\rm Spin}(D+~1)$. In other word, we get the
extra large symmetry here from the representation theory without
working out the symmetry generators explicitly.}
$$
{\ms H}=\bigoplus _{I=0}^{I_0(\kappa)}\,{\ms H}_I
$$ where ${\ms H}_I$ is a model for the irreducible $\mr{Spin}(D+1)$-representation
whose highest weight is $(I+|\mu|,|\mu|, \cdots, |\mu|, \mu)$;

3) The linear action in part 2) extends the manifest linear action
of $\mr{Spin}(D)$, and the finite dimensional complex vector space
${\ms H}_I$ in part 2) is the eigenspace of the hamiltonian $\hat h$
with eigenvalue $E_I$ in equation (\ref{spec}).
\end{Th}
Remark that there are continuous spectra unless $\kappa> 0$.

\subsection*{Acknowledgment}
I would like to thank Armen Nersessian for being interested in my
work on Kepler problems. I would also like to thank the conference
organizer for inviting me to write this report.

\end{document}